\documentclass[12pt]{article}

\usepackage[english]{babel}
\usepackage{graphicx}

   \usepackage{graphics,color,pstricks,pst-plot,pst-node,pst-coil}
   \usepackage[cp1251]{inputenc}

\begin{document}

\begin{center}
{\large \bf
High-Energy Standard Model from the Gauge Group Contraction
 }
\end{center}


\begin{center}
N.A. Gromov  \\
Institute of Physics and Mathematics,  \\
Komi Scientific Center, Ural Branch, Russian Academy of Sciences, \\
Syktyvkar, Russia \\
e-mail: gromov@ipm.komisc.ru
\end{center}

\begin{abstract}
The evolution of properties and interactions of elementary particles, beginning with the Planck scale of $10^{19}$ GeV. The description is based on the hypothesis that the 
high-temperature (high-energy) limit of the Standard Model is generated by the gauge group contraction. In the infinity-temperature limit, properties of particles fundamentally change: all particles lose their masses, and only massless neutral $Z$ bosons and $u$ quarks together with neutrinos and photons survive. Weak interactions become long-range ones and are generated by neutral currents. Quarks have only one color degree of freedom.
\end{abstract}


\section{Introduction}

Modern theory of the interaction of elementary particles, the Standard Model, includes the electroweak model, which combines electromagnetic and weak interactions, and quantum chromodynamics (QCD), which describes strong interactions. It describes the available ex\-peri\-men\-tal data quite well, and its adequacy has been convincingly confirmed by the recent discovery of the scalar Higgs boson in the experiments at the Large Hadron Collider. If one wants to investigate the properties and interactions of particles beyond the experimentally achieved energies, a possible way is to use the
 high-temperature (high-energy) limit of the Standard Model. This model is a gauge theory based on the $SU(3) \times SU(2) \times U(1) $ gauge group, which is a direct product of simple groups. Strong interactions of quarks are described by quantum chromodynamics with the $SU(3) $ gauge group and the characteristic temperature of $ 0.2 $ GeV. The electroweak model is based on the $ SU(2) \times U(1)$ gauge group, which is responsible for electroweak interactions with the characteristic temperature of $100 $ GeV. It follows from this observation that the gauge group of the theory of elementary particles becomes simpler with increasing temperature of the Universe. We assume that with further increase in the temperature, simplification of the gauge group of the Standard Model is described by its contraction.

The operation of contraction (or limiting transition) of groups \cite{IW-53}, which, in particular, transforms a simple group into a nonsemisimple one, is well known in physics. The notion of contraction was extended to algebraic structures, such as quantum groups and supergroups, and to fundamental representations of unitary groups \cite{Gr-12}. For a symmetric physical system, contraction of its symmetry group means a transition to the limiting state of the system. In the case of a complex physical system, which is the Standard Model, the study of its limiting states at any given limiting values of physical parameters allows a better insight into the behavior of the system as a whole. We will discuss the modified Standard Model with the contracted gauge group at the level of classical gauge fields.

In the broad sense of the word, deformation is the reverse of contraction. Nontrivial deformation of an algebraic structure means, generally speaking, its nonobvious generalization. A prominent recent example is quantum groups \cite{FRT}, i.e., such generalizations of Hopf algebras that are simultaneously noncommutative and noncocommutative, while previously Hopf algebras with only one of these properties were known. However, if a mathematical structure is first contracted, the initial structure can be reconstructed using deformation in the narrow sense, performed in the direction opposite to that of contraction.

We use this technique to reconstruct the evolution of elementary particles in the early Universe relying on the currently achieved knowledge. To this end, we consider the behavior of the Standard Model in the limit of the infinite temperature, which, according to our hypothesis, is generated by contraction of the $SU(2)$ and $SU(3) $ gauge groups \cite{Gr-2016}. Similar "infinitely" high temperatures could exist in the early Universe in the first instants after the Big Bang. It turns out that the gauge group contraction results in the Standard Model Lagrangian breaking down into a few terms with different powers of the zero-tending cont\-rac\-tion parameter $\epsilon\rightarrow 0$. Since the average energy (temperature) in the hot Universe is related to its age, then moving forward in time, i.e., in the direction opposite to the high-temperature contraction, we come to the conclusion that after the birth of the Universe the elementary particles and their interactions pass through a number of stages in their evolution from the limiting state with the infinite temperature to the state described by the Standard Model. 
These stages of quark--gluon plasma formation and reconstruction of electroweak and color symmetries differ by the powers of the cont\-rac\-tion parameter and consequently by the time of their origin. Based on the Standard Model contraction, we can classify these stages according to the "earlier--later" principle but cannot find the time elapsed after the birth of the Universe. To establish the absolute time, we use additional assumptions.

 \section{Electroweak model and its modification}

The Standard Model includes the following elementary particles:
gauge bosons
($
\mbox{photon}\; \gamma, \;
$
$
\mbox{charded and neutral weak} \; W^{\pm} \mbox{and} \; Z^{0} \; \mbox{bosons}, \;
\mbox{gluons}\; A^k,\; k=1,\ldots,8
$), 
a special particle
 $
\chi \; \mbox{(Higgs boson)},
$
and three generations of leptons 
$
\left(
\begin{array}{c}
	\nu_{e} \\
	e  
\end{array} \right), \;
$
$
\left(
\begin{array}{c}
	\nu_{\mu} \\
	\mu  
\end{array} \right), \;
$
$
\left(
\begin{array}{c}
	\nu_{\tau} \\
	\tau  
\end{array} \right)  \;
$
and quarks 
$
 \left(
\begin{array}{c}
	u \\
	d 
\end{array} \right), \;
$
$
\left(
\begin{array}{c}
	c \\
	s 
\end{array} \right), \;
$
$
\left(
\begin{array}{c}
	t \\ 
	b 
\end{array} \right),
$
described by vectors in  $\mathbf{C}_2 $.

We briefly describe the electroweak model according to \cite{R-99}. 
The model Lagrangian is equal to the sum of the boson, lepton and quark Lagrangians
$
L=L_B + L_L + L_Q
$
and is taken to be invariant under the action of the
 $ SU(2)\times U(1)$ gauge group in the space  $\mathbf{C}_2$.
 The boson sector $L_B=L_A + L_{\phi}$ consists of two parts, the gauge field Lagrangian 
$$ 
L_A=
 -\frac{1}{4}[(F_{\mu\nu}^1)^2 +(F_{\mu\nu}^2)^2+(F_{\mu\nu}^3)^2] -\frac{1}{4}(B_{\mu\nu})^2,
$$
\begin{equation}
F_{\mu\nu}=\partial_{\mu}A_{\nu}-\partial_{\nu}A_{\mu}+[A_\mu,A_\nu],\quad  B_{\mu\nu}=\partial_{\mu}B_{\nu}-\partial_{\nu}B_{\mu}
\label{eq2}
\end{equation}
and the matter field Lagrangian   
\begin{equation}
  L_{\phi}= \frac{1}{2}(D_\mu \phi)^{\dagger}D_\mu \phi -
  \frac{\lambda }{4}\left(\phi^{\dagger}\phi- v^2\right)^2, \quad
\phi= \left(
\begin{array}{c}
	\phi_1 \\
	\phi_2
\end{array} \right) \in \mathbf{C}_2.  
\label{eq3}
\end{equation}
The covariant derivatives are 
 \begin{equation}
D_\mu\phi=\partial_\mu\phi -ig\left(\sum_{k=1}^{3}T_kA^k_\mu \right)\phi-ig'YB_\mu\phi,
\label{eq4}
\end{equation}
where   $T_k=\frac{1}{2}\tau_k, k=1,2,3$ are the  $SU(2)$ generators, $\tau_{k}$ are the Pauli matrices,
 $Y=\frac{1}{2}{\bf 1}$ are the $U(1)$ generators, and $g$ and $g'$are the constants.

The fermion sector comprises the lepton $L_L$ and quark $L_Q$ Lagrangians.
The Lagrangian for the first generation is taken in the form
\begin{equation}
L_{L,e}=L_l^{\dagger}i\tilde{\tau}_{\mu}D_{\mu}L_l + e_r^{\dagger}i\tau_{\mu}D_{\mu}e_r -
h_e[e_r^{\dagger}(\phi^{\dagger}L_l) +(L_l^{\dagger}\phi)e_r],
\label{eq14}
\end{equation}
where
$
L_l= \left(
\begin{array}{c}
	\nu_l\\
	e_{l}
\end{array} \right) \in \mathbf{C}_2
$
is the  $SU(2)$ doublet,  $e_r $ is the  $SU(2)$ singlet, $h_e$ is the constant,
$\tau_{0}=\tilde{\tau_0}={\bf 1},$ $\tilde{\tau_k}=-\tau_k $, 
 $e_r, e_l, \nu_l $ are the two-component Lorentz spinors, and $D_{\mu} $ are the covariant derivatives of the lepton fields
$$
D_\mu L_l=\partial_\mu L_l -i\frac{g}{\sqrt{2}}\left(W_{\mu}^{+}T_{+} + W_{\mu}^{-}T_{-} \right)L_l-
 i\frac{g}{\cos \theta_w}Z_\mu\left( T_3 -Q\sin^2 \theta_w  \right)L_l -ieA_\mu Q L_l,
$$
\begin{equation}
D_{\mu}e_r = \partial_\mu e_r -ig'QA_\mu e_r \cos \theta_w +ig'QZ_\mu e_r \sin \theta_w,
\label{eq5-10}
\end{equation}
where 
$T_{\pm}=T_1\pm iT_2 $,
 $Q =Y+T_3$ is the generator of the $U(1)_{em}$ electromagnetic subgroup,
 $Y=\frac{1}{2}{\bf 1}$ is the hypercharge,
$ e=gg'(g^2+g'^2)^{-\frac{1}{2}} \;$ is the electron charge, and
$ \sin \theta_w=eg^{-1}.$

The quark Lagrangian is built in a similar way  
$$
L_Q=Q_l^{\dagger}i\tilde{\tau}_{\mu}D_{\mu}Q_l + 
u_r^{\dagger}i\tau_{\mu}D_{\mu}u_r +
d_r^{\dagger}i\tau_{\mu}D_{\mu}d_r -
$$
\begin{equation}
-h_d[d_r^{\dagger}(\phi^{\dagger}Q_l) +(Q_l^{\dagger}\phi)d_r]
-h_u[u_r^{\dagger}(\tilde{\phi}^{\dagger}Q_l) +(Q_l^{\dagger}\tilde{\phi})u_r],
\label{4}
\end{equation}
where the left quark fields make up the  $SU(2)$ doublet
$
Q_l= \left(
\begin{array}{c}
	u_l\\
	d_{l}
\end{array} \right)\in \mathbf{C}_2,
$
the right fields   $u_r, d_r $ are  $SU(2)$ singlets,
$\tilde{\phi}_i=\epsilon_{ik}\bar{\phi}_k, \epsilon_{00}=1, \epsilon_{ii}=-1$ 
make up a conjugate representation of the  $SU(2)$ group, and  $h_u, h_d$ are constants. 
All fields  $u_l, d_l, u_r, d_r $ are two-component Lorentz spinors. The covariant derivatives of the quark fields are
$$
D_{\mu}Q_l=\left(\partial_{\mu}- ig\sum_{k=1}^{3}\frac{\tau_k}{2}A^k_\mu -ig'\frac{1}{6}B_\mu  \right)Q_l,
$$
\begin{equation}
D_{\mu}u_r=\left(\partial_{\mu}-ig'\frac{2}{3}B_\mu  \right) u_r,\quad
D_{\mu}d_r=\left(\partial_{\mu}+ig'\frac{1}{3}B_\mu  \right) d_r.
\label{eq14-3Q}
\end{equation}

The  Lagrangians of the next lepton and quark generations are built similarly.
The total lepton and quark Lagrangians are obtained by summing over all generations. We confine ourselves to the consideration of the only the first generations.

The action of the contracted $SU(2;\epsilon)$ group in the fundamental representation space 
 $\mathbf{C}_2(\epsilon)$ is defined by the relation 
\begin{equation}
\left( \begin{array}{c}
	z'_1 \\
{\epsilon}	z'_2
\end{array} \right)
=\left(\begin{array}{cc}
	\alpha & {\epsilon}\beta   \\
-{\epsilon}\bar{\beta}	 & \bar{\alpha}
\end{array} \right)
\left( \begin{array}{c}
	z_1 \\
{\epsilon}	z_2
\end{array} \right), \quad
|\alpha|^2+{\epsilon^2}|\beta|^2=1
\label{5}
\end{equation}
as the contraction parameter tends to zero, 
 ${\epsilon} \rightarrow 0 $.
The replacement  $\beta \rightarrow {\epsilon} \beta $
induces transformation of the gauge fields, 
$
W_{\mu}^{\pm} \rightarrow {\epsilon}W_{\mu}^{\pm}, \; Z_{\mu} \rightarrow Z_{\mu},\; A_{\mu} \rightarrow A_{\mu}, \;
$
and the replacement  
$z_2 \rightarrow {\epsilon}z_2 $ 
induces transformation of the lepton and quark fields,
$
e_{l} \rightarrow {\epsilon} e_{l},  \;  d_{l} \rightarrow {\epsilon} d_{l}, \;
\nu_l \rightarrow \nu_l, \;  	u_l \rightarrow u_l.\;
$
In the spontaneous symmetry breaking mechanism, one of the main states  
$
  \phi^{vac}=\left(\begin{array}{c}
	0  \\
	v
\end{array} \right), \;  A_\mu^k=B_\mu=0, \;
$
 of the Lagrangian $L_B$ is taken as the model vacuum, and then  low excitations of the fields
$ v+\chi(x) $ with respect to this vacuum are considered. 
Therefore, the Higgs boson field  $ \chi $, constant  $ v $, and   $ v $-dependent particle masses $m_p$ 
are multiplied by the contraction parameter:
$
 	\chi  \rightarrow {\epsilon} \chi,  \;  v \rightarrow {\epsilon} v, \; 	 m_p \rightarrow {\epsilon} m_p, \; p={\chi}, W, Z, e, u, d.\;
$

After these transformations, the Lagrangian of the electroweak model takes the form
 \begin{equation}
 L({\epsilon})= L_{\infty} - {\epsilon} \, m_u(u_r^{\dagger}u_{l} + u_{l}^{\dagger}u_r)+
 {\epsilon^2} L_{2} + {\epsilon^3} L_{3}  + {\epsilon^4} L_{4}.
\label{9}
\end{equation}
When  ${\epsilon} \rightarrow 0 $, the terms with higher powers of   ${\epsilon} $ 
make a smaller contribution   compared to the terms with lower powers. 
Thus, the electroweak model demonstrates five stages of the behavior as ${\epsilon} \rightarrow 0 $, 
which differ by the powers of the contraction parameter.
  
\section{QCD with contracted gauge group}

The QCD  $SU(3)$ gauge group acts in the space  $\mathbf{C}_3$ of the color quark states 
$
q=(q_1, q_2, q_3)^t = (q_R, q_G, q_B)^t \in \mathbf{C}_3, \;
$
where  $q(x)$ are the quarks,   $q=u, d, s, c, b, t $; and  $R$, $G$ and $B$ are the color degrees of freedom.
 The QCD Lagrangian is taken in the form 
\begin{equation}
{\cal L} =\sum_q \left(\bar{q}^i(i\gamma^\mu)(D_\mu)_{ij}q^j -m_q \bar{q}^iq_i\right)
-\frac{1}{4}\sum_{\alpha=1}^8
F_{\mu\nu}^\alpha F^{\mu\nu\, \alpha},
 \label{q1}
\end{equation}
where  $D_\mu q$ are the covariant derivatives of the quark fields 
 \begin{equation}    
D_\mu q=  
\left(\partial_{\mu}-ig_s\left(\frac{\lambda^\alpha}{2}\right)A^\alpha_{\mu}\right)q,
 \label{q19}
\end{equation}
$g_s$ is the strong coupling constant,  $t^a=\lambda^a/2$ are the $SU(3)$ group generators, $\lambda^a $ are the Gell--Mann
matrices, and the gluon strength tensor has the a standard form  
 \begin{equation}   
F_{\mu\nu}^\alpha=\partial_{\mu} A_\nu^\alpha-\partial_{\nu} A_\mu^\alpha+
g_sf^{\alpha\beta\gamma}A_\mu^\beta A_\nu^\gamma.
 \label{q29}
\end{equation}

The contracted   $SU(3;{\epsilon})$ group is defined by the action 
\begin{equation}
q'({\epsilon})=\left(\begin{array}{c}
 q'_{1}\\
{\epsilon} q'_{2} \\
{\epsilon^2} q'_{3}
 \end{array}
 \right)=
\left(\begin{array}{ccc}
 u_{11}  &{\epsilon} u_{12} &{\epsilon^2} u_{13} \\
 {\epsilon} u_{21} & u_{22} & {\epsilon} u_{23} \\
 {\epsilon^2} u_{31} & {\epsilon} u_{32} & u_{33}
 \end{array}
 \right)
 \left(\begin{array}{c}
 q_{1}\\
{\epsilon} q_{2} \\
{\epsilon^2} q_{3}
 \end{array}
 \right)=U({\epsilon} )q({\epsilon} )
 \label{10}
\end{equation}
in the color space  $\mathbf{C}_3({\epsilon})$ as  ${\epsilon} \rightarrow 0 $.
Therefore, the quark and gluon fields transform as follows:
$
q_1 \rightarrow q_1, \;
q_2\rightarrow {\epsilon}  q_2,\; q_3\rightarrow {\epsilon^2} q_3,
$
$
A_\mu^{GR}\rightarrow {\epsilon} A_\mu^{GR},\;
A_\mu^{BG}\rightarrow {\epsilon} A_\mu^{BG},\;
A_\mu^{BR}\rightarrow {\epsilon^2} A_\mu^{BR},\;
$
while the diagonal gluon fields do not change: 
$
A_\mu^{RR}\rightarrow A_\mu^{RR},\;
A_\mu^{GG}\rightarrow A_\mu^{GG},\;
A_\mu^{BB}\rightarrow A_\mu^{BB}.
$
After these substitutions, we obtain the quark part of the QCD Lagrangian in the form  
$$
{\cal L}_q({\epsilon} )= \sum_q \Biggl\{ i\bar{q}_1\gamma^\mu\partial_{\mu} q_1 -m_q\left|q_1\right|^2
+\frac{g_s}{2} \left|q_1\right|^2 \gamma^\mu A_\mu^{RR}+
$$
$$
+{\epsilon^2} \biggl\{
i\bar{q}_2\gamma^\mu \partial_{\mu} q_2 -m_q\left|q_2\right|^2  +
\frac{g_s}{2}\biggl(\left|q_2\right|^2 \gamma^\mu A_\mu^{GG}+
q_1\bar{q}_2\gamma^\mu A_\mu^{GR}+ \bar{q}_1q_2\gamma^\mu \bar{A}_\mu^{GR}\biggr) \biggr\} +
$$
$$
+{\epsilon^4} \biggl[
i\bar{q}_3\gamma^\mu \partial_{\mu} q_3 -m_q\left|q_3\right|^2  +
\frac{g_s}{2}\biggl(\left|q_3\right|^2 \gamma^\mu A_\mu^{BB}+
q_1\bar{q}_3\gamma^\mu A_\mu^{BR}+ \bar{q}_1q_3\gamma^\mu \bar{A}_\mu^{BR}+
$$
\begin{equation}
+ q_2\bar{q}_3\gamma^\mu A_\mu^{BG}+ \bar{q}_2q_3\gamma^\mu \bar{A}_\mu^{BG}
\biggr) \biggr] \Biggr\}
 = L_q^{\infty} + {\epsilon^2} L_q^{(2)} + {\epsilon^4} L_q^{(4)}.
 \label{q10}
\end{equation}
The gluon part  $L_{gl}=-\frac{1}{4}F_{\mu\nu}^\alpha F^{\mu\nu\, \alpha}$
of the Lagrangian is very cumbersome, and we omit it. 
The transformed QCD Lagrangian takes the form 
\begin{equation}
{\cal L}({\epsilon})=
L^{\infty} + {\epsilon^2} L^{(2)} + {\epsilon^4} L^{(4)}+ {\epsilon^6} L^{(6)}+ {\epsilon^8} L^{(8)},
 \label{11}
\end{equation}
with the known expressions for each  $L^{(k)}$.

\section{Estimation of boundary values}

Combining (\ref{9}) and (\ref{11}), we write the Standard Model Lagrangian in the form
\begin{equation}
{\cal L}_{SM}({\epsilon})=
{\cal L}_{\infty} + {\epsilon} {\cal L}_{1} + {\epsilon^2} {\cal L}_{2} + 
{\epsilon^3}{\cal L}_{3} + {\epsilon^4} {\cal L}_{4}+ {\epsilon^6} {\cal L}_{6}+ {\epsilon^8} {\cal L}_{8}.
 \label{12}
\end{equation}
According to the adopted hypothesis, the contraction parameter is monotonic function of temperature
${\epsilon}(T) \rightarrow 0 $ as $T \rightarrow \infty $.
Under modern concepts of the origin of our Universe \cite{GoR-11}, very high temperatures exist
in it at the first stages of the Big Bang immediately after inflation (Fig. 1).

\begin{figure}[h]
\begin{center}
\includegraphics [height=0.25 \textheight,trim=0 0 0 0, clip]{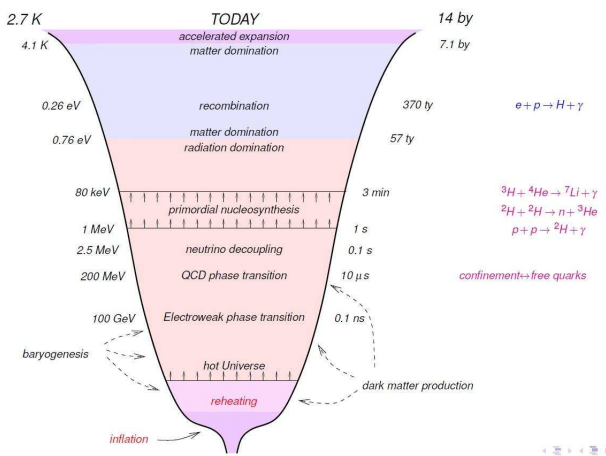}
\end{center}
\caption{History of the Universe \cite{GoR-11} $\,$ ($ 1 eV=10^{4} K$).}
\end{figure}

Contraction of the gauge group of the Standard Model makes it possible to chronologically order its development stages
bu does not allow the absolute dates to be established. To estimate them, we use the fact that the electroweak epoch begins at the temperature
 $T_4=100$ GeV ($1$ GeV $=10^{13} K$), and the QCD epoch begins at   $T_8=0.2$ GeV.
In other words, we assume that it is at these temperatures that complete reconstraction of the electroweak model,
whose Lagrangian involves terms proportional to ${\epsilon^4}$, and reconstruction of QCD with the minimal
Lagrangians terms of the order  $\epsilon^{8}$ take place. 

Let $\Delta$ be the cutoff level for    
${\epsilon^k}, \; k=1,2,4,6,8$, i.e., when  ${\epsilon^k} < \Delta$, all Lagrangian terms proportional to ${\epsilon^k}$ 
are negligibly small.  Finally, we assume that the contraction parameter  is proportional to the inverse temperature 
\begin{equation}
\epsilon(T)=\frac{A}{T}, \quad A=const.
 \label{q14}
\end{equation}
From the QCD equation     
${\epsilon^8}(T_8)=A^8T_8^{-8}= \Delta $
we obtain  $A=T_8\Delta^{1/8}=0,2\Delta^{1/8}$ GeV.
From a similar electroweak model equation, we find the cutoff level 
${\Delta} = (T_8E_4^{-1})^8=(0,2\cdot10^{-2})^8\approx {10^{-22}}$.
From the equation for the  $k$th pover
${\epsilon^k}(T_k)=A^kT_k^{-k}=\Delta $
we have
$
T_k= T_8\Delta^{\frac{k-8}{8k}}\approx 10^{\frac{88-15k}{4k}}
$ 
and easily find boundary values (GeV) 
\begin{equation}
T_1=10^{18},\;\; T_2=10^7,\;\; T_3=10^3,\;\; T_4=10^2,\;\; T_6=1,\; T_8=2\cdot10^{-1}
 \label{14}
\end{equation}
The estimate of the "infinite" $\,$ temperature  
$T_1\approx 10^{18} $ GeV is comparable with the Planck energy
$\approx 10^{19}$ GeV, at which the gravitation effect becomes substantial.
Thus, the resulting evolution of elementary particles does not go beyond the problems described by 
electroweak and strong interactions.
 
\section{Evolution of particles} %

We may draw some conclusions even at the level of the classical fields.
At the "infinite"$\,$ temperature  (${\epsilon} =0, $ $ T> 10^{18}$ GeV) 
the Lagrangian of the electroweak model has the form 
$$ 
L_{\infty}= - \frac{1}{4}{\cal Z}_{\mu\nu}^2 - \frac{1}{4}{\cal F}_{\mu\nu}^2 +
\nu_l^{\dagger}i\tilde{\tau}_{\mu}\partial_{\mu}\nu_l
+u_{l}^{\dagger}i\tilde{\tau}_{\mu}\partial_{\mu}u_{l}  +
$$
\begin{equation}
+e_r^{\dagger}i\tau_{\mu}\partial_{\mu}e_r +
d_r^{\dagger}i\tau_{\mu}\partial_{\mu}d_r
+u_r^{\dagger}i\tau_{\mu}\partial_{\mu}u_r + L_{\infty}^{int}(A_{\mu},Z_{\mu}),
 \label{14w}
\end{equation}
i.e., the electroweak model contains only massless particles: photons  $A_{\mu}$ and neutral  $Z_{\mu}$ bosons, 
left quarks  $u_{l}$ and neutrinos $\nu_l$, right electrons  $e_r $ and quarks  $ u_r, d_r $.
Masslessness has a simple physical explanation: the temperature is so high that 
the particle mass is negligibly small in comparison to the kinetic energy.
Electroweak interactions become long-range ones, since they are transferred by massless 
 $Z$ bosons and photons.  

It follows from the explicit expression for the interacting Lagrangian 
$$
L_{\infty}^{int}(A_{\mu},Z_{\mu})= {\frac{g}{2\cos \theta_w} \nu_l^{\dagger}\tilde{\tau}_{\mu}Z_{\mu}\nu_l}
+\frac{2e}{3}u_{l}^{\dagger}\tilde{\tau}_{\mu}A_{\mu}u_{l}+ \frac{g}{\cos \theta_w}\left(\frac{1}{2}-\frac{2}{3}\sin^2\theta_w\right) u_{l}^{\dagger}\tilde{\tau}_{\mu}Z_{\mu}u_{l} +
$$
 $$
 +
g'\sin \theta_w e_r^{\dagger}\tau_{\mu}Z_{\mu}e_r 
- g'\cos \theta_w e_r^{\dagger}\tau_{\mu}A_{\mu}e_r
-\frac{1}{3}g'\cos\theta_w d_r^{\dagger}\tau_{\mu}A_{\mu}d_r + \frac{1}{3}g'\sin\theta_w d_r^{\dagger}\tau_{\mu}Z_{\mu}d_r +
$$
\begin{equation}
 + \frac{2}{3}g'\cos\theta_w u_r^{\dagger}\tau_{\mu}A_{\mu}u_r
-\frac{2}{3}g'\sin\theta_w u_r^{\dagger}\tau_{\mu}Z_{\mu}u_r
 \label{15w}
\end{equation}
that particles of different kind do not interact with one another, e.g., neutrinos interact with each other through neutral currents.
This looks like a sort of stratification of the electroweak model with particles of the same kind in each layer,

From the limiting QCD Lagrangian
$$
{\cal L}^{\infty}=L_q^{\infty}+L_{gl}^{\infty}=
\sum_q i\bar{q}_1\gamma^\mu\partial_{\mu} q_1 -m_q\left|q_1\right|^2
+\frac{g_s}{2} \left|q_1\right|^2 \gamma^\mu A_\mu^{RR}-
$$
\begin{equation}
-\frac{1}{4}\left(F_{\mu\nu}^{RR}\right)^2
-\frac{1}{4}\left(F_{\mu\nu}^{GG}\right)^2
-\frac{1}{4}F_{\mu\nu}^{RR}F_{\mu\nu}^{GG}
 \label{16w}
\end{equation}
we conclude that at the "infinite" temperature only terms for one color component of massless quarks survive, i.e., quarks become monochromatic. 
Terms describing interaction of these components with $R$ gluons also persist. 
Thus, there is stratification in the QCD sector as well.

In the temperature interval $10^{18}\; \mbox{GeV} \geq T > 10^7 $ GeV the mass terms of the $u$ quark in the total Lagrangian 
  ${\cal L}({\epsilon}) $ are proportional to the contraction parameter,
$
{\epsilon} \, m_u(u_r^{\dagger}u_{l} + u_{l}^{\dagger}u_r).
$  
 This also applies to the $c$ and $t$ quarks. Therefore, the $u, c$,  and $t$ quarks are the first to restore their mass in the process of evolution of the Universe.

In the interval $10^7\; \mbox{GeV} \geq T > 10^3 $ GeV mass terms of the electron and the $d$ quark are proportional to the square of the contraction parameter, 
$
{{\epsilon^2}}\, \left[m_e(e_r^{\dagger}e_l + e_l^{\dagger} e_r)+
 m_d(d_r^{\dagger}d_{l} + d_{l}^{\dagger}d_r)\right].
$
 The same is true for the $\mu $ and $\tau $ leptons and the $s $ and $b$ quarks. These particles become massive at the second stage. Quarks acquire the second color degree of freedom. The main part of the electroweak and color interactions is restored in that epoch.

At temperatures $10^3\; \mbox{GeV} \geq T > 10^2 $ GeV the Lagrangian has only one term 
$L_3=gW_\mu^+W_\mu^-\chi$
 proportional to ${\epsilon^3}$ that describes interaction of the Higgs boson with charged $W$ bosons. In the next epoch
$  10^{2}\; \mbox{GeV} \geq T> 1 $ GeV the Higgs boson  $ \chi$  and charged $W$ bosons are the last to acquire mass. The electroweak model is ultimately restored. Quarks acquire the third color degree of freedom.

At temperatures $  1\,$GeV $\geq T > 0,2$ GeV there are all color interactions except 
$L_{gl}^{(8)}= 
-\frac{g_s^2}{4} \Bigl(A_{\mu}^4A_{\nu}^5 - A_{\mu}^5A_{\nu}^4\Bigr)^2. $ 
Finally, at $ T \leq 0,2 $  GeV the Standard Model is entirely restored.

\section{Conclusions}

The high-temperature limit of the Standard Model obtained from first principles of the gauge theory upon gauge group contraction is investigated. The mathe-matical contraction parameter is taken to be inversely proportional to the temperature of the Universe, and its zero limit corresponds to the "infinite" temperature of the order of the Planck energy of 1019 GeV. During the evolution of the Universe, the Standard Model passes through several stages that differ by the powers of the contraction parameter. The limiting temperatures between them are found by introducing the cutoff level $ \Delta $ with consideration of typical QCD and electroweak model energies. On the basis of expansions (\ref{9}), (\ref{11}), and (\ref{12}), intermediate Lagrangians $ {\cal L}_{k} $ are built for each stage of the Standard Model development, which allows conclusions to be drawn about interactions and properties of particles in each of the epochs.

The resulting scheme of evolution of elementary particles from the first instants after the Big Bang does not contradict the history of the Universe devised from other reasoning ([GoR-11], Fig. 1), according to which QCD phase transitions occur later than electroweak phase transitions. In addition, it provides grounding for more detailed analysis of the genesis stages of leptons and quark--gluon plasma in view of the fact that the terms $L^{(6)}_{gl}$ and $L^{(8)}_{gl}$ in the QCD gluon Lagrangian $L_{gl} $ \cite{Gr-2020} are negligibly small at temperatures of 0.2 to 100 GeV, and in the temperature interval of 100 to 1000 GeV only the interaction of the Higgs boson with charged $ W$  bosons is restored.

\end{document}